\documentclass[journal]{IEEEtran}

\usepackage{pifont}
\usepackage{eurosym}
\usepackage{amsmath}
\usepackage{amssymb}
\usepackage{mathtools}
\usepackage[utf8]{inputenc}
\usepackage[vietnamese,english]{babel}
\usepackage{cite}
\usepackage{slashbox}

\ifCLASSINFOpdf
\usepackage[pdftex]{graphicx}
\else
\fi

\usepackage{amsmath}
\usepackage{soul}

\usepackage{url}

\usepackage{makecell}
\usepackage[table, dvipsnames]{xcolor}
\usepackage[acronym,toc,shortcuts]{glossaries}
\usepackage{csquotes}

\usepackage{tabularx}

\ifCLASSOPTIONcompsoc
\usepackage[caption=false,font=normalsize,labelfon
t=sf,textfont=sf]{subfig}
\else
\usepackage[caption=false,font=footnotesize]{subfi
g}
\fi

\newcolumntype{s}{>{\columncolor[HTML]{C9DFEC}} l}

\newcolumntype{x}{>{\columncolor[HTML]{E5E4E2}} l}

\arrayrulecolor[HTML]{DB5800}

\newacronym{2D}{2D}{two dimensional}
\newacronym{3D}{3D}{three dimensional}
\newacronym{5G}{5G}{Fifth Generation}
\newacronym{5G PPP}{5G PPP}{5G Infrastructure Public Private Partnership}
\newacronym{3GPP}{3GPP}{3rd Generation Partnership Project}
\newacronym{4D}{4D}{four dimensional}
\newacronym{AAA}{AAA}{Authentication, Authorization and Accounting}
\newacronym{ABS}{ABS}{Aerial Base Station}
\newacronym{ABSF}{ABSF}{Almost-Blank Subframe}
\newacronym{AES}{AES}{Advanced Encryption Standard }
\newacronym{AI}{AI}{Artificial Intelligence}
\newacronym{AMC}{AMC}{Adaptive Modulation and Coding}
\newacronym{AP}{AP}{access point}
\newacronym{API}{API}{Application Programming Interface}
\newacronym{APN}{APN}{Access Point Name}
\newacronym{AR}{AR}{Augmented Reality}
\newacronym{AWGN}{AWGN}{additive white Gaussian noise}
\newacronym{BBU}{BBU}{Baseband Unit}
\newacronym{BCN}{BCN}{Blockchain Network}
\newacronym{BE}{BE}{best-effort}
\newacronym{BET}{BET}{Blind Equal Throughput}
\newacronym{BLAST}{BLAST}{Bell Laboratories Layered Space-Time}
\newacronym{BLER}{BLER}{Block Error Rate}
\newacronym{BS}{BS}{Base Station}
\newacronym{BTP}{BTP}{Backhaul Transport Provider}
\newacronym{BTS}{BTS}{Base Transceiver Station}
\newacronym{CA}{CA}{carrier aggregation}
\newacronym{MBAR}{MBAR}{Mobile Backhaul Aggregation Router}
\newacronym{CAPEX}{CapEx}{capital expenditure}
\newacronym{CDF}{CDF}{Cumulative Distribution Function}
\newacronym{CELL-ID}{CELL-ID}{cell identification ID}
\newacronym{CIO}{CIO}{cell individual offset}
\newacronym{CDN}{CDN}{Content Delivery Network}
\newacronym{CN}{CN}{core network}
\newacronym{CP}{CP}{Control Plane}
\newacronym{CPU}{CPU}{central processing unit}
\newacronym{CoMP}{CoMP}{Coordinated Multipoint}
\newacronym{CSR}{CSR}{Cell Site Router}
\newacronym{CQI}{CQI}{Channel Quality Indicator}
\newacronym{C-RAN}{C-RAN}{Cloud RAN}
\newacronym{CS}{CS}{central scheduler}
\newacronym{CSI}{CSI}{channel state information}
\newacronym{CRE}{CRE}{cell range expansion}
\newacronym{D2D}{D2D}{Device-to-Device}
\newacronym{DLT}{DLT}{Distributed Ledger Technology}
\newacronym{DFT}{DFT}{discrete Fourier transform}
\newacronym{DSL}{DSL}{Digital subscriber line}
\newacronym{EARFCN}{EARFCN}{E-UTRA Absolute Radio Frequency Channel Number}
\newacronym{EC}{EC}{European Commission}
\newacronym{eICIC}{eICIC}{enhanced inter-cell interference cancellation}
\newacronym{eMBB}{eMBB}{enhanced Mobile Broadband}
\newacronym{eNodeB}{eNodeB}{Evolved Node B}
\newacronym{EPC}{EPC}{Evolved Packet Core}
\newacronym{EPS}{EPS}{Evolved Packet System}
\newacronym{ETSI}{ETSI}{European Telecommunications Standards Institute}
\newacronym{E-UTRAN}{E-UTRAN}{Evolved Universal Terrestrial Radio Access Network}
\newacronym{FANET}{FANET}{Fly Ad Hoc Network}
\newacronym{FDMA}{FDMA}{frequency division multiple access}
\newacronym{FFT}{FFT}{fast Fourier transform}
\newacronym{FSO}{FSO}{Free-Space Optical Communication}
\newacronym{FTP}{FTP}{File Transfer Protocol}
\newacronym{FU}{FU}{Frame Usage}
\newacronym{GTP}{GTP}{GPRS Tunneling Protocol}
\newacronym{GGSN}{GGSN}{Gateway GPRS Support Node}
\newacronym{GPS}{GPS}{global positioning system}
\newacronym{GRA}{GRA}{Grey relational analysis}
\newacronym{GSM}{GSM}{Global System for Mobile Communications}
\newacronym{GEO}{GEO}{Geosynchronous}
\newacronym{GIS}{GIS}{Geographical Information Systems}
\newacronym{GTP-U}{GTP-U}{GPRS Tunneling Protocol-User Plane}
\newacronym{HAPS}{HAPS}{High Altitude Platform Station(s)}
\newacronym{HDFS}{HDFS}{Hadoop Distributed File System}
\newacronym{HetNet}{HetNet}{Heterogeneous Network}
\newacronym{HiveQL}{HiveQL}{Hive Query language}
\newacronym{HD}{HD}{High Definition}
\newacronym{HEO}{HEO}{High Earth Orbit}
\newacronym{HO}{HO}{handover}
\newacronym{HARQ}{HARQ}{Hybrid automatic repeat request}
\newacronym{HS-DSCH}{HS-DSCH}{High Speed Downlink Shared Channel}
\newacronym{HSS}{HSS}{Home Subscriber Station}
\newacronym{HTS}{HTS}{High Throughput Satellite}
\newacronym{HTTP}{HTTP}{Hypertext Transfer Protocol}
\newacronym{IAB}{IAB}{Integrated Access and Backhaul}
\newacronym{ICIC}{ICIC}{inter-cell interference cancellation}
\newacronym{ICN}{ICN}{information-centric network}
\newacronym{IEEE}{IEEE}{Institute of Electrical and Electronics Engineers}
\newacronym{IMEI}{IMEI}{International Mobile Station Equipment Identity}
\newacronym{IMSI}{IMSI}{International Mobile Subscriber Identity}
\newacronym{IMS}{IMS}{IP Multimedia Subsystem}
\newacronym{IMT-A}{IMT-A}{International Mobile Telecommunications - Advanced}
\newacronym{ITU}{ITU}{International Telecommunication Union}
\newacronym{IP}{IP}{Internet Protocol}
\newacronym{IPsec}{IPsec}{Internet Protocol Security}
\newacronym{IoT}{IoT}{Internet of Things}
\newacronym{ISAC}{ISAC}{Integrated Sensing and Communication}
\newacronym{JSON}{JSON}{JavaScript Object Notation}
\newacronym{KPI}{KPI}{key performance indicator}
\newacronym{LAC}{LAC}{location area code}
\newacronym{LEO}{LEO}{Low Earth Orbit}
\newacronym{LoS}{LoS}{Line-of-Sight}
\newacronym{LPWAN}{LPWAN}{Low Power Wide Area Network}
\newacronym{LTE}{LTE}{Long Term Evolution}
\newacronym{LTE-A}{LTE-A}{Long Term Evolution Advanced}
\newacronym{mmWave}{mmWave}{millimeter wave}
\newacronym{MAC}{MAC}{Medium Access Control}
\newacronym{MADM}{MADM}{Multiple Attribute Decision Making}
\newacronym{MANET}{MANET}{Mobile Ad Hoc Network}
\newacronym{MBH}{MBH}{Mobile Backhaul}
\newacronym{MCS}{MCS}{Modulation Coding Scheme}
\newacronym{MEO}{MEO}{Medium Earth Orbit}
\newacronym{MEW}{MEW}{multiplicative exponent weighting}
\newacronym{MIMO}{MIMO}{multiple-input multiple-output}
\newacronym{ML}{ML}{Machine Learning}
\newacronym{MME}{MME}{Mobility Management Entity}
\newacronym{mMTC}{mMTC}{massive Machine Type Communications}
\newacronym{MMF}{MMF}{max-min fairness}
\newacronym{MMSE}{MMSE}{minimum mean square error}
\newacronym{MPLS}{MPLS}{Multiprotocol Label Switching}
\newacronym{MSISDN}{MSISDN}{Mobile Station International Subscriber Directory Number}
\newacronym{MSP}{MSP}{Mobile Service Provider}
\newacronym{MT}{MT}{Maximum Throughput}
\newacronym{NAS}{NAS}{Non Access Stratum}
\newacronym{NE}{NE}{Nash Equilibrium}
\newacronym{NLP}{NLP}{Natural Language Processing}
\newacronym{NR}{NR}{New Radio}
\newacronym{NTN}{NTN}{Non-Terrestrial Network}
\newacronym{NFV}{NFV}{Network Functions Virtualization}
\newacronym{NoSQL}{NoSQL}{Not Only SQL}
\newacronym{OAM}{OAM}{Operation, Administration and Management}
\newacronym{OFDM}{OFDM}{orthogonal frequency division multiplexing}
\newacronym{OFDMA}{OFDMA}{orthogonal frequency division multiple access}
\newacronym{ONF}{ONF}{open networking foundation}
\newacronym{ONOS}{ONOS}{Open Network Operating System}
\newacronym{OPEX}{OpEx}{operating expenditure}
\newacronym{OS}{OS}{operating system}
\newacronym{OTT}{OTT}{over-the-top}
\newacronym{PCI}{PCI}{Physical Cell Identity}
\newacronym{PCRF}{PCRF}{Policy and Charging Rules Function}
\newacronym{PDF}{PDF}{Probability Distribution Function}
\newacronym{PDN}{PDN}{packet data network}
\newacronym{PDCP}{PDCP}{Packet Data Convergence Control}
\newacronym{PDSCH}{PDSCH}{Physical Downlink Shared Channel}
\newacronym{PDU}{PDU}{Protocol Data Unit}
\newacronym{PF}{PF}{Proportional Fair}
\newacronym{PGW}{P-GW}{Packet Data Gateway}
\newacronym{PHY}{PHY}{physical layer}
\newacronym{PoC}{PoC}{Proof-of-Concept}
\newacronym{PPP}{PPP}{{P}oisson point process}
\newacronym{PTP}{PTP}{Precision Time Protocol}
\newacronym{QoE}{QoE}{quality-of-experience}
\newacronym{QoS}{QoS}{quality-of-service}
\newacronym{QCI}{QCI}{QoS Class Identifier}
\newacronym{PSC}{PSC}{Primary Scrambling Code}
\newacronym{PSD}{PSD}{power spectral density}
\newacronym{RACH}{RACH}{random access channel}
\newacronym{RAN}{RAN}{Radio Access Network}
\newacronym{RAT}{RAT}{Radio Access Technology}
\newacronym{RB}{RB}{Resource Block}
\newacronym{RE}{RE}{range extension}
\newacronym{RF}{RF}{radio frequency}
\newacronym{RG}{RG}{rate guarantee}
\newacronym{RLC}{RLC}{Radio Link Controller}
\newacronym{RNC}{RNC}{Radio Network Controller}
\newacronym{RR}{RR}{Round Robin}
\newacronym{RRC}{RRC}{Radio Resource Control}
\newacronym{RRH}{RRH}{remote radio head}
\newacronym{RRU}{RRU}{Remote Radio Unit}
\newacronym{RRM}{RRM}{radio resource management}
\newacronym{RSI}{RSI}{RACH Root Sequence Index}
\newacronym{RSS}{RSS}{received signal strength}
\newacronym{RSSI}{RSSI}{received signal strength indicator}
\newacronym{RSRP}{RSRP}{reference signal received power}
\newacronym{RTT}{RTT}{Round Trip Time}
\newacronym{SAC}{SAC}{service area code}
\newacronym{SANET}{SANET}{Sea Ad Hoc Network}
\newacronym{SAW}{SAW}{simple additive weighting}
\newacronym{SC-FDMA}{SC-FDMA}{single carrier frequency division multiple access}
\newacronym{SCN}{SCN}{small cell network}
\newacronym{SCTP}{SCTP}{Stream Control Transmission Protocol}
\newacronym{SDN}{SDN}{Software-Defined Networking}
\newacronym{SDO}{SDO}{Standard Development Organization}
\newacronym{SDMN}{SDMN}{Software Defined Mobile Network}
\newacronym{SDU}{SDU}{Service Data Unit}
\newacronym{SecGW}{SecGW}{Security Gateway}
\newacronym{SGSN}{SGCN}{Serving GPRS Support Node}
\newacronym{SGW}{S-GW}{Serving Gateway}
\newacronym{SHARING}{SHARING}{Self-organized Heterogeneous Advanced RadIo Networks Generation}
\newacronym{SNR}{SNR}{signal-to-noise ratio}
\newacronym{SINR}{SINR}{signal-to-interference-plus-noise ratio}
\newacronym{SISO}{SISO}{single-input single-output}
\newacronym{SSID}{SSID}{Service Set Identification}
\newacronym{ST}{ST}{Standart Multi-User TOPSIS}
\newacronym{STBCs}{STBCs}{space-time block codes}
\newacronym{SyncE}{SyncE}{Synchronous Ethernet}
\newacronym{TB}{TB}{Transport Block}
\newacronym{TBS}{TBS}{Transport Block Size}
\newacronym{TCP}{TCP}{Transport Control Protocol}
\newacronym{TDMA}{TDMA}{Time Division Multiple Access}
\newacronym{TEID}{TEID}{tunnel endpoint identifier}
\newacronym{TOPSIS}{TOPSIS}{Total Order Preference By Similarity to the Ideal Solution}
\newacronym{TTI}{TTI}{transmission time interval}
\newacronym{UAV}{UAV}{Uncrewed Aerial Vehicle}
\newacronym{UAV-BS}{UAV-BS}{Unmanned Aerial Vehicles-Base Station}
\newacronym{UARN}{UARN}{UAV-aided relay network}
\newacronym{UDP}{UDP}{User Datagram Protocol}
\newacronym{UE}{UE}{user equipment}
\newacronym{UL}{UL}{Uplink}
\newacronym{UQD}{UQD}{UAV-BS QoS Determination}
\newacronym{UP}{UP}{User Plane}
\newacronym{UMTS}{UMTS}{Universal Mobile Telecommunications Service} 
\newacronym{URLLC}{URLLC}{Ultra-reliable low latency communications}
\newacronym{VANET}{VANET}{Vehicular Ad Hoc Network}
\newacronym{VoIP}{VoIP}{voice over IP}
\newacronym{VPN}{VPN}{virtual private network}
\newacronym{VR}{VR}{Virtual Reality}
\newacronym{VSAT}{VSAT}{Very Small Aperture Terminal}
\newacronym{W-CDMA}{W-CDMA}{Wideband Code Division Multiple Access}
\newacronym{WiFi}{WiFi}{Wireless Fidelity}
\newacronym{Wi-Fi}{Wi-Fi}{Wireless Fidelity}
\newacronym{WiMAX}{WiMAX}{Worldwide Interoperability for Microwave Access}
\newacronym{WLAN}{WLAN}{Wireless Local Area Network}
\newacronym{WMC}{WMC}{weighted Markov chain}
\newacronym{ZF}{ZF}{zero-forcing}
\newacronym{MNO}{MNO}{Mobile Network Operator}
\newacronym{RES}{RES}{Renewable Energy Sources}
\newacronym{SON}{SON}{Self Organizing Network}
\newacronym{ANR}{ANR}{Automatic Neighbor Relation}
\newacronym{MRO}{MRO}{Mobility Robustness Optimizer}
\newacronym{MLB}{MLB}{Mobility Load Balancing}
\newacronym{CQO}{CQO}{cell quality offset}
\newacronym{BESS}{BESS}{Battery Energy Storage System}
\newacronym{GBS}{GBS}{Ground Base Station}
\newacronym{PV}{PV}{Photovoltaic}
\newacronym{DMC}{DMC}{Disaster Management Center}
\newacronym{SEMA}{SEMA}{Sustainable Energy Management Algorithm}
\newacronym{DEMA}{DEMA}{Disaster Energy Management Algorithm}
\newacronym{EV}{EV}{Electric Vehicle}
\newacronym{MDRU}{MDRU}{Movable and Deployable Resource Unit}
\newacronym{V2G}{V2G}{Vehicle to Grid}
\newacronym{G2V}{G2V}{Grid to Vehicle}
\newacronym{VHetNet}{VHetNet}{Vertical Heterogeneous Network}
\newacronym{IRS}{IRS}{Intelligent Reflective Surfaces}
\newacronym{mMIMO}{mMIMO}{massive multiple-input multiple-output}
\newacronym{WPAN}{WPAN}{Wireless Personal Area Network}
\newacronym{CR}{CR}{Charging Rate}
\newacronym{V2L}{V2L}{Vehicle to Load}
\newacronym{GSP}{GSP}{Grid Selling Price}
\newacronym{GBP}{GBP}{Grid Buying Price}
\newacronym{DSO}{DSO}{Distribution System Operator}
\newacronym{SOC}{SOC}{State of Charge}
\newacronym{THz}{THz}{Terahertz}

\newacronym{ICT}{ICT}{Information and Communication Technology}

\renewcommand{\arraystretch}{2}

\usepackage[subtle]{savetrees}

\makeatletter
\patchcmd{\@maketitle}
  {\addvspace{0.5\baselineskip}\egroup}
  {\addvspace{-1\baselineskip}\egroup}
  {}
  {}
\makeatother

\begin{document}

\bstctlcite{IEEEexample:BSTcontrol}

% Do not put math or special symbols in the title.
%\title{Increasing Disaster Resiliency and Sustainability of Integrated Space-Air-Ground-Sea  Networks with Renewable Energy Solutions: A Use Case after the Turkiye Earthquake}
\title{On-Demand HAPS-Assisted Communication System for Public Safety in Emergency and Disaster Response}

\author{Bilal Karaman, Ilhan Basturk, ~\IEEEmembership{Senior Member,~IEEE},  Ferdi Kara, ~\IEEEmembership{Senior Member,~IEEE}, Engin Zeydan, ~\IEEEmembership{Senior Member,~IEEE}, \\  Esra Aycan Beyazit, ~\IEEEmembership{Member,~IEEE,}, Sezai Taskin, Emil Bj\"ornson, ~\IEEEmembership{Fellow,~IEEE} and Halim Yanikomeroglu, ~\IEEEmembership{Fellow,~IEEE}.  
        % <-this % stops a space
% \thanks{C.E. Kement, and H. Yanikomeroglu are with the Department of Systems and Computer Engineering, Carleton University, Ottawa, ON, K1S 5B6 Canada, e-mails: \{cihankement, halim\}@sce.carleton.ca.} 
% \thanks{F. Kara is with the Department of Computer Engineering, Zonguldak Bulent Ecevit University, Turkey and is also with the Department of Systems and Computer Engineering, Carleton University, Ottawa, ON, K1S 5B6 Canada, e-mail: f.kara@beun.edu.tr.}
% \thanks{W. Jaafar is with the Department of Software and IT Engineering, \'{E}cole de Technologie Sup\'{e}rieure, Montreal, QC, Canada, e-mail: wael.jaafar@etsmtl.ca.}
% \thanks{G. Senarath, N. D. Dào, and P. Zhu are with Huawei Technologies Co., Ltd, Canada e-mail: \{gamini.senarath, ngoc.dao, peiying.zhu@huawei.com.}
% \thanks{$^*$F. Kara and W. Jaafar contributed equally 
% to the paper.}
% <-this % stops a space
%\thanks{Manuscript received April 19, 2005; revised August 26, 2015.}
\thanks{The work of F. Kara and E. Bj\"ornson is supported by the SweWIN Vinnova Competence Center.}
\thanks{B. Karaman, I. Basturk and S. Taskin are with Manisa Celal Bayar University, Turkiye. B. Karaman is also with  Department of Systems and Computer Engineering, Carleton University, Ottawa, ON, K1S 5B6 Canada, emails: \{bilal.karaman, ilhan.basturk, sezai.taskin\}@cbu.edu.tr.}
\thanks{E. Zeydan is with Centre Tecnològic de Telecomunicacions de Catalunya (CTTC), Barcelona, Spain, 08860, email: ezeydan@cttc.es.} 
\thanks{F. Kara and E. Bj\"ornson are with Department of Computer Science, KTH Royal Institute of Technology, Stockholm, Sweden, 16440. F. Kara is also with Department of Computer Engineering, Zonguldak Bulent Ecevit University, Zonguldak Turkiye, 67100, e-mails: \{ferdi, emilbjo\} @kth.se.}
\thanks{E. A. Beyazıt is with IDLab Research Group, University of Antwerp - IMEC, Belgium, e-mails: esra.aycanbeyazit@imec.be.}
\thanks{H. Yanikomeroglu is with Department of Systems and Computer Engineering, Carleton University, Ottawa, ON, K1S 5B6 Canada, e-mail: halim@sce.carleton.ca.}
}

\renewcommand{\baselinestretch}{1.1}
\selectfont %now it works

% The paper headers
\markboth{IEEE Communications Magazine}%
{Karaman \MakeLowercase{\textit{et al.}}: Enhancing Resiliency of Integrated}
% The only time the second header will appear is for the odd numbered pages
% after the title page when using the twoside option.
% 
% *** Note that you probably will NOT want to include the author's ***
% *** name in the headers of peer review papers.                   ***
% You can use \ifCLASSOPTIONpeerreview for conditional compilation here if
% you desire.

% If you want to put a publisher's ID mark on the page you can do it like
% this:
%\IEEEpubid{0000--0000/00\$00.00~\copyright~2015 IEEE}
% Remember, if you use this you must call \IEEEpubidadjcol in the second
% column for its text to clear the IEEEpubid mark.

% use for special paper notices
%\IEEEspecialpapernotice{(Invited Paper)}

% make the title area
\maketitle

\begin{abstract}

Natural disasters often disrupt communication networks and severely hamper emergency response and disaster management. Existing solutions, such as portable communication units and cloud-based network architectures, have improved disaster resilience but fall short if both the Radio Access Network (RAN) and backhaul infrastructure become inoperable. To address these challenges, we propose a demand-driven communication system supported by High Altitude Platform Stations (HAPS) to restore communication in an affected area and enable effective disaster relief. The proposed emergency response network is a promising solution as it provides a rapidly deployable, resilient communications infrastructure. The proposed HAPS-based communication can play a crucial role not only in ensuring connectivity for mobile users but also in restoring backhaul connections when terrestrial networks fail. As a bridge between the disaster management center and the affected areas, it can facilitate the exchange of information in real time, collect data from the affected regions, and relay crucial updates to emergency responders. Enhancing situational awareness, coordination between relief agencies, and ensuring efficient resource allocation can significantly strengthen disaster response capabilities. In this paper, simulations show that HAPS with hybrid optical/THz links boosts backhaul capacity and resilience, even in harsh conditions. HAPS-enabled RAN in S- and Ka-bands ensures reliable communication for first responders and disaster-affected populations. This paper also explores the integration of HAPS into emergency communication frameworks and standards, as it has the potential to improve network resilience and support effective disaster management.

%\textcolor{red}{An alternative abstract is given below. Any of them can be used.}
%, by developing practical strategies and implementing resilient and sustainable network architectures

\end{abstract}

% \begin{IEEEkeywords}
% 6G networks, disaster resiliency, disaster recovery.
% \end{IEEEkeywords}

% Note that keywords are not normally used for peerreview papers.

%Our aim is to contribute to the development of more robust and sustainable disaster response frameworks, by developing practical strategies and implementing resilient and sustainable network architectures. 

% For peer review papers, you can put extra information on the cover
% page as needed:
% \ifCLASSOPTIONpeerreview
% \begin{center} \bfseries EDICS Category: 3-BBND \end{center}
% \fi
%
% For peerreview papers, this IEEEtran command inserts a page break and
% creates the second title. It will be ignored for other modes.
\IEEEpeerreviewmaketitle

%\tableofcontents

\section{Introduction}

%\hl{Ferdi: I wrote the following by using some AI tools to create examples of ongoing studies. İlhan: Bende HAPS ile birleştirerek asagıdaki gibi birsey olusturdum Ferdi Hocam. Sizin kısımları commente aldım. Bir akısa bakalım hangisi uygunsa kullanırız.}

Natural disasters, such as earthquakes, hurricanes, floods, and wildfires, have long demonstrated their catastrophic impact on communities worldwide. These events often cause massive loss of life, destruction of property, and extensive disruption to essential infrastructure, particularly communication networks. For example, the $2004$ Indian Ocean tsunami devastated coastal regions in Southeast Asia, claiming over $230,000$ lives and displacing millions. Similarly, the $2010$ earthquake in Haiti left more than $160,000$ people dead and led to widespread devastation, exposing significant weaknesses in the country’s communication systems. In the United States, Hurricane Katrina in $2005$ caused severe damage to New Orleans, disrupting communication networks and hindering emergency response efforts. These events have all underscored the vulnerability of existing communication infrastructures in times of disaster, making it clear that effective communication systems are critical for timely response and recovery. The recent earthquakes in Türkiye in February $2023$ are a stark reminder of these challenges. With magnitudes of $7.7$ and $7.6$, these earthquakes devastated entire cities in southern Türkiye, claiming over $50,000$ lives and displacing millions of people. The destruction of critical infrastructure, especially communication networks, made it difficult for rescue teams and survivors to make contact immediately after the disaster. The collapse of ground-based communication systems severely hampered rescue efforts, delayed coordination, and exacerbated the crisis, highlighting the urgent need for more resilient and rapidly deployable communication solutions in disaster scenarios \cite{karaman2024enhancing}.
\begin{figure*}[htp!]
    \centering
    \includegraphics[width=0.9\linewidth]{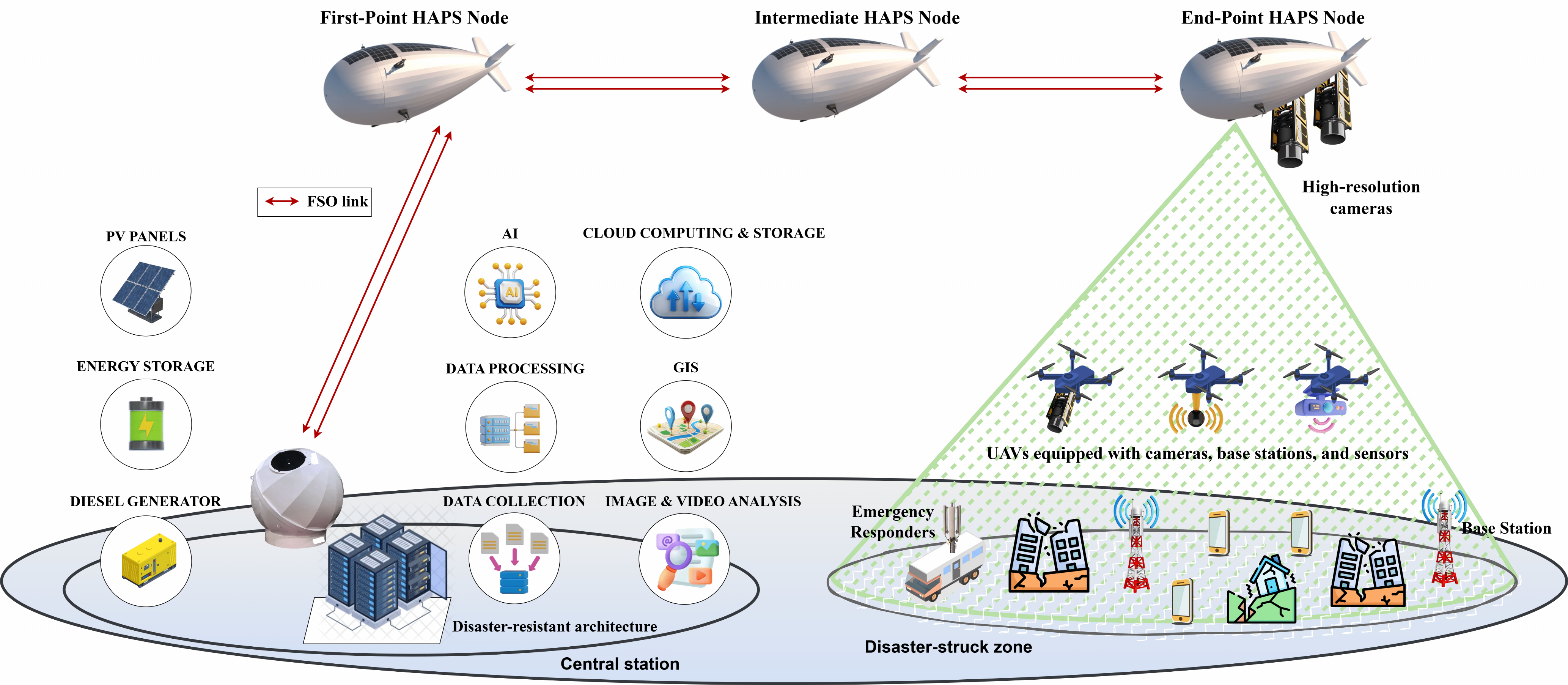}
    \caption{Illustration of the proposed on-demand HAPS-assisted private network.}
    \label{fig:system_model}
    \vspace{-.5cm}
\end{figure*}

In response to these challenges, technical solutions have been developed to improve communication in emergencies. Public safety agencies around the world are increasingly focused on ensuring that communications systems remain operational during disasters. In the United States, for example, the Federal Communications Commission (FCC) has introduced the Mandatory Disaster Response Initiative (MDRI) to improve the resilience of wireless networks. This initiative includes measures such as bilateral roaming agreements and mutual aid between service providers to reduce network outages and speed up recovery. Portable communication systems such as Cells on Wheels (COWs) have also been deployed in Australia to ensure connectivity in areas affected by bushfires and flooding. These systems are rapidly deployed in disaster areas to maintain communication between emergency responders and the affected population when traditional infrastructure is damaged. The authors in \cite{ma2022joint} propose a resilient, multi-layered network architecture that integrates the existing infrastructure with new layers, including cloud processing. The new architecture aims to improve the scalability and resilience of communication networks in the event of a disaster. Cloud computing is integrated to provide flexible and scalable \ac{ICT} services and reduce dependency on physical infrastructure. Geo-redundancy and load-balancing techniques are also implemented to manage network traffic and ensure high availability. In \cite{ma2022joint}, two new optimal integer linear program models are proposed that exploit the intrinsic interplay between data and service evacuations and can efficiently use the early warning time before cloud data centers are affected.  

However, while these solutions help to a degree, they still have limitations, particularly in the face of widespread disasters that render existing communication infrastructures unusable. Following the 2023 Türkiye earthquakes, for instance, large swathes of the affected regions were without power for extended periods, further complicating efforts to restore telecommunications. These challenges highlight the need for even more innovative and adaptable solutions to bridge communication gaps in such critical times.  

One potential solution is the use of High Altitude Platform Stations (HAPS) \cite{kurt2021vision}. In disaster scenarios, it is studied for different aims like improving access link or backhaul link  \cite{karaman2024enhancing, 10097717} or providing aerial computing services for the terrestrial
Internet of Things (IoT) devices \cite{9714482}. Operating in the stratosphere, HAPS offer several advantages that make them suitable for disaster response. Their ability to cover vast areas, rapid deployment, and minimal reliance on ground-based infrastructure make them a powerful tool for providing on-demand communication in disaster-stricken regions. HAPS can establish reliable communication channels quickly, ensuring that emergency responders and affected populations can communicate with each other and coordinate relief efforts. This can significantly enhance situational awareness, improve coordination between agencies, and facilitate more efficient resource allocation, ultimately saving lives and reducing the impact of the disaster. This paper explores the potential of a HAPS-assisted communication system tailored specifically for public safety during emergency and disaster response. By addressing the communication gaps exposed during past disasters, including the Türkiye earthquakes, HAPS could play a crucial role in enhancing the resilience of disaster management frameworks, ensuring that timely and effective responses are possible even in the most challenging conditions.

\section{Proposed Architecture}
%\hl{@Bilal, cizim, @Ferdi yazi}

%\subsection{Proposed Scenario}

%(\hl{this is an example figure I created roughly, we can work on that together with Bilal})

%\textbf{Architecture \& Optimizations:} 

The proposed private on-demand network consists of a central station, intermediate HAPS nodes, and an end-point HAPS node, as shown in Fig. \ref{fig:system_model}. It has its isolated backhaul through HAPS nodes using FSO links and HAPS-assisted Radio Access Network (RAN) that operates on both existing cellular frequencies and Ka-band. The central station must be located in a secure region, as it is responsible for controlling and deploying the private network when necessary. This central station acts as a disaster management and control center, integrating high computing capacity for efficient operations. The control station should be centrally located (e.g., in the geographical center of the country) to minimize the distance to disaster-affected areas. It includes a ground station for communication with the first HAPS intermediate node. All collected data is processed at this central station to facilitate disaster response through various management and planning strategies, which will be detailed in the following sections. Although the proposed on-demand private network is primarily for public safety, it also has gateways to connect commercial communication networks that can serve as a parallel core network to support communication in the disaster area. In addition to the technical advantages, locating a disaster management center in a region that is not affected by the disaster also has psychological advantages (e.g., it is common practice to appoint a temporary governor or decision-making staff in the affected areas to prevent people from acting emotionally in such cases). The central station can also include energy technology enablers such as \ac{PV} panels, battery storage systems, and diesel generators to ensure power continuity.

The proposed on-demand emergency network has intermediate HAPS nodes that are responsible for backhaul communication between the ground station and the affected region. %These backhaul links between intermediate HAPS-nodes and from/to the ground station are comprised of Free-Space-Optics (FSO) links. FSO communication has the capacity of replacing fiber-optic wired backhaul links as long as a line-of-sight (LoS) link is secured. Since the HAPS-nodes are operated nearly 20 km above the earth, a direct LoS can be guaranteed without any natural event obstacles (such as clouds, fog, rain) since these events occur at much lower altitude.
Depending on the distance to the ground station, several nodes may be required. These HAPS nodes can be designed to fly independently or be transported as cargo to other locations. Recent advances in the field of HAPS show that these nodes can be deployed and transferred to a certain distance in just a few hours. These flying objects are equipped with solar panels to provide the energy needed to relay the communication load from the affected area to the central station and vice versa. 

The end-point HAPS node in the proposed system serves multiple functions in restoring communication and collecting data in disaster-affected regions. It assists terrestrial \glspl{BS} in re-establishing (damaged) backhaul connections and provides backhaul or uplink communication for \glspl{UAV}. Some \glspl{UAV} function as \glspl{BS} when terrestrial infrastructure is disrupted, while others, equipped with sensors and cameras, support disaster assessment through real-time data collection.
%In addition to FSO links, THz communication may also be needed in this case since the FSO links can be affected by weather events while THz has much more resiliency against these events. These FSO/THz link can also be used to coordinate UAVs,. 
In addition, the end HAPS node itself can communicate with users on the ground if required.  %Particularly, emergency-related data (e.g., shelter, aid) can be broadcasted with a direct-to-device service and critical communication with first responders can be guaranteed in Ka band. 
Finally, the end HAPS node may also carry some high-resolution cameras that can be used to capture footprints (in clear sky) from the region, just as we now use satellite imagery services for disaster management assessment and planning.

% Bilal: CS ile ilgili figure ENgin hocamiz ile gorusup, revize edip, section 3te paylasalim. ENABLER'larin figure olsun. Bu sekle Ka ve S band ekle. UAV BASE Station yap ve altina Userler at beam ile.  En asagiya Users ve S band ile Hapstan in. Vsat anten koyup ona da haps tan Ka band ile in. Simulasyonda Fig 2 ye refere et.

\section{Technology Enabler for Proposed Scenario: \\ Private Network}

\subsection{Over-the-Air Communication}

% \hl{@Ferdi}

\begin{figure}[]
    \centering
    \includegraphics[width=0.9\linewidth]{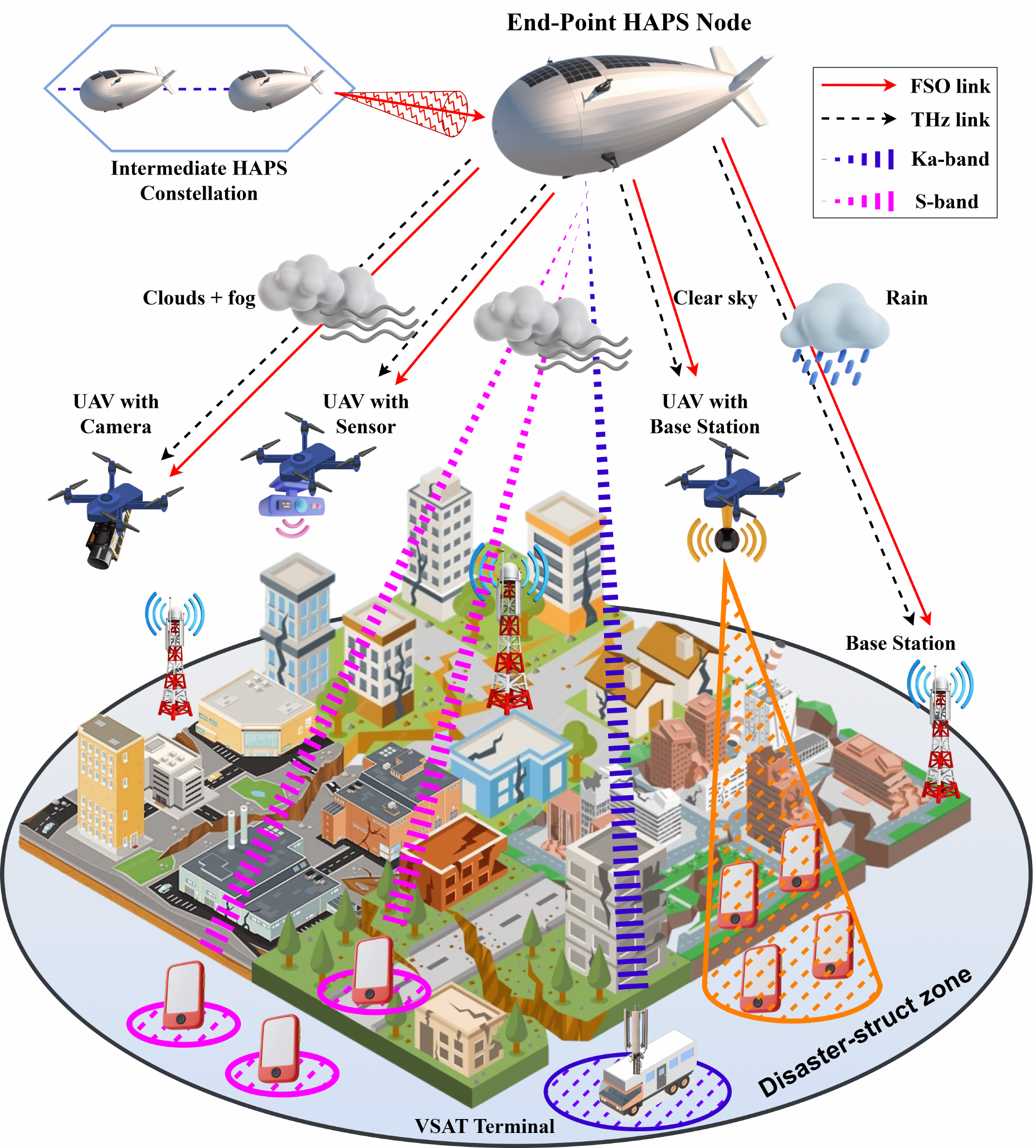}
    \caption{Disaster area communication network.}
    \label{fig:Fig2b}
    \vspace{-8pt}
\end{figure}

\textit{Inter-HAPS Free-Space Optics (FSO) Communication: Backhaul:} 
Disaster assessment and management require a vast amount of data sharing between the first responders and the operation planning center. This data includes the collected images from the area, command communication for the rescue operation, and planning for first aid such as shelter, food, etc. These inter-HAPS backhaul links can be established with FSO communication from the ground station to the end-point HAPS node, as shown in Fig. \ref{fig:Fig2b}. FSO communication has the capacity to replace fiber-optic wired backhaul links as long as a Line-of-Sight (LoS) link is secured. Since the HAPS nodes are operated nearly 20 km above the earth, a direct LoS can be guaranteed without any weather events (such as clouds, fog, or rain) since these events occur at much lower altitudes than the stratosphere.

\textit{Hybrid FSO/THz Communication: Fronthaul/Backhaul for Terrestrial BSs and UAVs:}
Terrestrial telecommunication infrastructure can be severely affected by disasters. In most cases, the terrestrial BS might be completely unavailable due to physical damage, power outage, etc. On the other hand, the fronthaul/backhaul links might be disrupted, although the BS itself can stay operational. The end-point HAPS node can assist terrestrial infrastructure in these scenarios. A backhaul link can be established from HAPS to operational terrestrial \glspl{BS} through a hybrid FSO/\ac{THz} communication. Unlike inter-HAPS communication, THz communication may also be needed in this case since the FSO links can be affected by weather events, while THz has much more resiliency against these events. Therefore, a hybrid FSO/THz communication link seems an optimal solution, as shown in Fig. \ref{fig:Fig2b}.  However, as mentioned above, the terrestrial BS might be completely nonoperational. In this scenario, temporary (several) UAV-BS can be transferred to those regions to replace the terrestrial BSs. The established FSO/THz communication backhaul links to UAVs will be crucial to restore communication in these regions.

In addition to restoring communication, surveillance and data collection from the region have a crucial role in effective disaster response planning. In this context, UAVs equipped with cameras and sensors play a vital role in disaster assessment and rescue operations. However, collecting data from these high-resolution cameras and transferring them to the operation center has challenges when the network is already facing disruptions and congestion. Therefore, the hybrid FSO/THz communication from end-point HAPS to surveillance UAVs (as shown in Fig. \ref{fig:Fig2b}) can guarantee uninterrupted data collection to have effective disaster response planning for authorities.

\textit{Direct Access to Ground Users: RAN:}
The end-point HAPS node is also capable of providing direct access to ground users. To eliminate the need for additional devices, direct communication with mobile handheld devices should be established in the S-band (the same frequencies with existing cellular networks) so they can connect to HAPS just as terrestrial BS. This capability plays a critical role in delivering essential information, such as shelter locations, first aid availability, and food distribution points, to affected populations across all regions. Since this service is expected to require low data rates, it can be effectively supported by the end-point HAPS. Meanwhile, uninterrupted communication between the coordination center and first responders is crucial for effective disaster management. Under current conditions, first responders typically communicate with the command center via push-to-talk radios or satellite links. However, these systems have several limitations, including limited range and service disruptions. To overcome these challenges, direct access from HAPS can ensure high-data-rate, uninterrupted communication, provided that first responder vehicles are equipped with \glspl{VSAT} in the Ka-band. The Radio Frequency (RF) communication strategy in both scenarios aligns with ITU's frequency allocations for HAPS as defined in \cite{ITU_WRC23}.

\subsection{The Role of Mobile Applications, AI and Data Processing in the Central Station}
%Social media, AI, vision, LLM,  Data Center, Core Network
%5G core, 

%\hl{@Engin: Data centerda toplanan verilerle neler yapilabilir? Ornegin Blockchain kaldirilavilir. Cunku biz datalari toplayip merkezi bir yapida toplamaya calisiyoruz. Ornegin blockchain ise tam tersine datalarin distributed yapida islenmesini oneriyor vb. Bu sectionda Architecture degisebilir. }

%\textbf{AI at data centers:} 
%have been proved valuable in post-disaster and crisis management, e.g. recovering and consolidating information, searching and rescuing with limited human interaction, and post-disaster assessment . 
%/\ac{ML}

The data collected in the zone affected by the disaster can be analyzed in the central station, which is not affected by the disaster.  While traditional data analytics can process and interpret this information, \ac{AI} techniques can provide a more advanced, adaptive and automated approach that enables a faster and more efficient response to disasters, especially in situations where predefined human-generated models are too rigid or cannot capture the complexity and unpredictability of disaster events. Therefore, \ac{AI} techniques can play a critical role in reducing disaster risk by predicting extreme events and creating hazard maps, enabling real-time event detection and situational awareness, providing decision support for emergency responders, and automating emergency networks \cite{ETSI_TR_102_445_2023}. This improves root cause analysis, predictive maintenance, and network infrastructure optimization. \ac{AI} algorithms can serve various purposes in disaster management. First, they optimize resource distribution by analyzing the needs of affected areas, predicting demand, and determining the most efficient delivery routes \cite{chamola2020disaster}. Second, they assess the impact of disasters by analyzing satellite imagery, drone footage, and other visual data to automatically evaluate damage. Third, \ac{AI} can process vast amounts of data from social media to provide real-time insights that help locate survivors, identify urgent needs, and track the spread of both accurate information and misinformation during post-disaster recovery.  On the other hand, virtualization and cloud computing can enable emergency services to process and analyze large amounts of data in real time. These technologies can offer flexibility, fast recovery, and better security through firewalls and microservices. Cloud-based architectures that support resilient and fault-tolerant emergency networks can ensure continuity during disasters. 

%Big data analytics and predictive modeling can be used for resource allocation and long-term planning.  Analysis using big data tools helps understand the impact of the disaster, optimize resource allocation, and make informed decisions for long-term recovery efforts. %\cite{swaminathan2018big}. 

%\cite{jamali2019social,sadhukhan2018producing}.  

%In reference \cite{yulianto2018rapid}, the authors have introduced RAPID TIMEER, which is a self-sufficient system that can document and share information (using text, images, and voice) without relying on power or telecommunication infrastructures. This system utilizes client-server communication and has a coverage range of up to 12km. 

%\textbf{Positioning systems:} 

\ac{GIS} can serve as effective tools for the visualization, analysis, and interpretation of disaster-related data and support all phases of disaster management \cite{van2002remote}. First, \ac{GIS} improves risk assessment and preparedness by identifying areas at risk through the analysis of geographic and demographic data. By evaluating factors such as population density, infrastructure distribution, and historical disaster trends, \ac{GIS} enables better disaster planning and mitigation strategies. Second, \ac{GIS} improves the situational awareness of disasters in real time by integrating data from multiple sources, including weather reports, satellite imagery and sensor networks. This dynamic mapping capability allows emergency responders to monitor unfolding events, predict affected areas, and adjust response strategies accordingly. Third, \ac{GIS} plays a crucial role in optimizing resources and logistics by identifying effective paths and distribution centers. By analyzing transportation networks and accessibility constraints, GIS can aid disaster management teams in allocating resources effectively and ensure the timely delivery of emergency aid. Fourth, GIS facilitates evacuation planning and execution by designing optimal evacuation routes based on terrain analysis, road conditions, and population movement patterns. This helps authorities to implement safer and more effective evacuation plans to reduce casualties and congestion during emergencies.  Finally, GIS supports post-disaster recovery and public communication by mapping the affected areas, assessing the damage, and identifying the critical infrastructure that needs to be restored. In addition, GIS improves public awareness by presenting clear visualizations of risks, safety measures, and recovery efforts, helping communities make informed decisions in disaster situations.

%\ac{GIS} technologies can provide powerful tools for visualizing, analyzing, and interpreting data related to disaster risks, vulnerabilities, and resources \cite{van2002remote}. \ac{GIS} helps identify vulnerable areas by analyzing geographic and demographic data, enabling better planning and preparedness. During disasters, \ac{GIS} integrates real-time data like weather reports and satellite imagery to provide up-to-date situational awareness. \ac{GIS} optimizes resource allocation by identifying the most efficient routes and locations for aid distribution. By analyzing transportation networks and population density, \ac{GIS} helps design effective evacuation routes and plans. After disasters, \ac{GIS} assesses damage and informs recovery efforts by mapping affected areas and identifying critical infrastructure. \ac{GIS} tools also enhance public communication by providing clear visual information on risks, safety measures, and recovery efforts.

%\textbf{Mobile Applications:} 

%Mobile applications can be a powerful tool for disseminating real-time information, evacuation routes and safety guidelines during an earthquake and recovery efforts. 

Mobile applications serve as an effective tool for providing real-time updates, guiding evacuation routes, and delivering safety guidelines during earthquakes and recovery efforts. Over-the-Top (OTT) apps, including apps for voice and data communication, video distribution, and conferencing, are widely used by emergency services. Some apps are directly connected to the Public Switched Telephone Network for emergency calls. Many OTT apps rely on cloud-based solutions and virtualization to ensure scalability and security. These apps support session recording and multi-party connectivity, which is essential for public safety answering points. Mobile OTT applications can provide users with critical information, enabling them to make informed decisions and take necessary actions to stay safe during and after an earthquake. For instance, applications such as \enquote{MyShake} and \enquote{ShakeAlert} utilize smartphone sensors to detect earthquakes and issue early warnings upon detection. Beyond early alerts, these applications can also offer guidance on emergency procedures, evacuation paths, and a \enquote{Safe Corner} property to aid users identify the safest sheltering spots. Additionally, they provide comprehensive resources on earthquake preparedness, safety measures, and recovery efforts, along with interactive maps displaying the locations of nearby earthquake early warning sensors. Additionally, technologies such as \ac{AR} and \ac{VR} can contribute to disaster preparedness by simulating crisis scenarios and providing hands-on training for emergency responders. These immersive tools enhance crisis management skills, improving real-world disaster response capabilities.

%By providing users with access to important information, mobile OTT apps can help people make informed decisions and take appropriate actions to stay safe during and after an earthquake.  Apps such as  \enquote{MyShake} \enquote{ShakeAlert} use the sensors in the user's smartphone to detect earthquakes and warn the user at an early stage when an earthquake is detected. In addition to the early warning system, these apps also contain information on what to do in the event of an earthquake, evacuation routes, and a \enquote{Safe Corner} feature that helps users find the safest place to seek shelter in the event of an earthquake.  These apps also include information on earthquake safety, preparedness, and recovery, as well as a map showing the locations of nearby earthquake early warning sensors. 

%Some examples of successful mobile apps in this respect are the \enquote{MyShake} app\footnote{Online: https://myshake.berkeley.edu/, Available: February 2024.} and ShakeAlert\footnote{Online: https://www.shakealert.org/, Available: February 2024.}. 
\begin{table*}[h]
    \scriptsize
    \centering
    \caption{Simulation parameters used in evaluations for each setup.}
    \renewcommand{\arraystretch}{1.5} % Increases row height for readability
    \begin{tabular}{|s|x|s|x|s|x|s|x|}
        \hline
        \rowcolor{lightgray}
        \multicolumn{2}{|c|}{\textbf{FSO Link}} & \multicolumn{2}{c|}{\textbf{THz Link}} & \multicolumn{2}{c|}{\textbf{Ka-band Link}} & \multicolumn{2}{c|}{\textbf{S-band Link}} \\
        \hline
        \textbf{Parameter} & \textbf{Value} & \textbf{Parameter} & \textbf{Value} & \textbf{Parameter} & \textbf{Value} & \textbf{Parameter} & \textbf{Value} \\
        \hline
        Laser wavelength & 1550 nm & Frequency & 144 GHz & Frequency & 30 GHz & Frequency & 2.4 GHz \\
        \hline
        Transmit power & 17.5 dBm & Transmit power & 17.5 dBm & Transmit power & 43.2 dBm & Transmit power & 43.2 dBm \\
        \hline
        Tx and Rx optical efficiencies & 0.8 & Signal bandwidth & 30 GHz & Signal bandwidth & 400 MHz & Signal bandwidth & 100 MHz \\
        \hline
        Receiver telescope diameter & 80 mm & Tx antenna gain & 55 dBi & Tx antenna gain & 13.8 dBi & Tx antenna gain & 13.8 dBi \\
        \hline
        Tx and Rx pointing errors & 1 $\mu$rad & Rx antenna gain & 55 dBi & Rx antenna gain & 39.7 dBi & Rx antenna gain & 0 dBi \\
        \hline
        Full transmitting divergence angle & 15 $\mu$rad & Noise spectral density & - 174 dBm/Hz & Noise spectral density &  - 174 dBm/Hz & Noise spectral density & - 174 dBm/Hz \\
        \hline

    \end{tabular}
    \label{tab:sims}
\end{table*}

\section{Case Study}

In this section, we demonstrate the communication capabilities of the proposed on-demand HAPS network in a disaster-struck area. We show simulation results for each communication protocol discussed in the previous section. Overall simulation parameters for each setup are given in Table \ref{tab:sims}.

\begin{figure}[]
    \centering
    \includegraphics[width=0.9\linewidth]{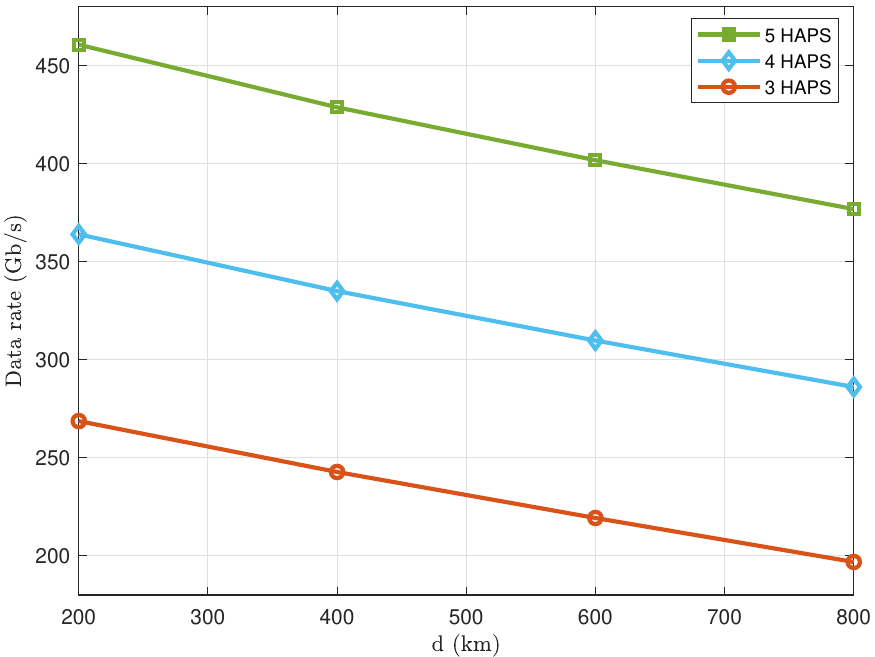}
    \caption{End-to-end backhaul capacity for the proposed on-demand HAPS network by using FSO communication. The x-axis shows the bird's-eye distance of the disaster region to the central station, and the y-axis shows the achievable backhaul capacity with different intermediate HAPS nodes. }
    \label{fig:Backhaul}
    \vspace{-12pt}
\end{figure}

To demonstrate the backhaul capabilities of HAPS networks in Fig. \ref{fig:system_model}, we obtained the simulation results for the data rates achieved at the HAPS endpoint as shown in Fig.~\ref{fig:Backhaul}. This figure illustrates the data rates at the HAPS endpoint as a function of the distance between the central station and the region potentially affected by the disaster, taking into account a different number of intermediate HAPS nodes. In our simulations, the HAPS nodes are located at the first-point, intermediate-point, and end-point at an altitude of $20$ km. Since there is no rain or clouds in the stratosphere under normal atmospheric conditions, we use FSO for data transmission between the central station and the end-point HAPS node. Additionally, the transmitter and receiver antenna gains, along with atmospheric attenuations due to Mie scattering and geometrical scattering, are modeled as proposed in \cite{liang2022link}, while cloud, fog, and rain attenuations are based on \cite{alzenad2018fso}. The remaining communication parameters can be found in Table \ref{tab:sims}. As expected, with more intermediate nodes, the backhaul capacity increases. Nevertheless, even in the worst-case scenario with $3$ HAPS (including the end-point HAPS node) in total $800$ km distance to the disaster area, $200$ Gbps data can be transferred in both directions which is quite enough for a backhaul link. This amount can increase up to $450$ Gbps in a shorter distance with more intermediate HAPS nodes.

% \hl{@Bilal, can you quantify?}

\begin{figure}[]
    \centering
    \includegraphics[width=0.9\linewidth]{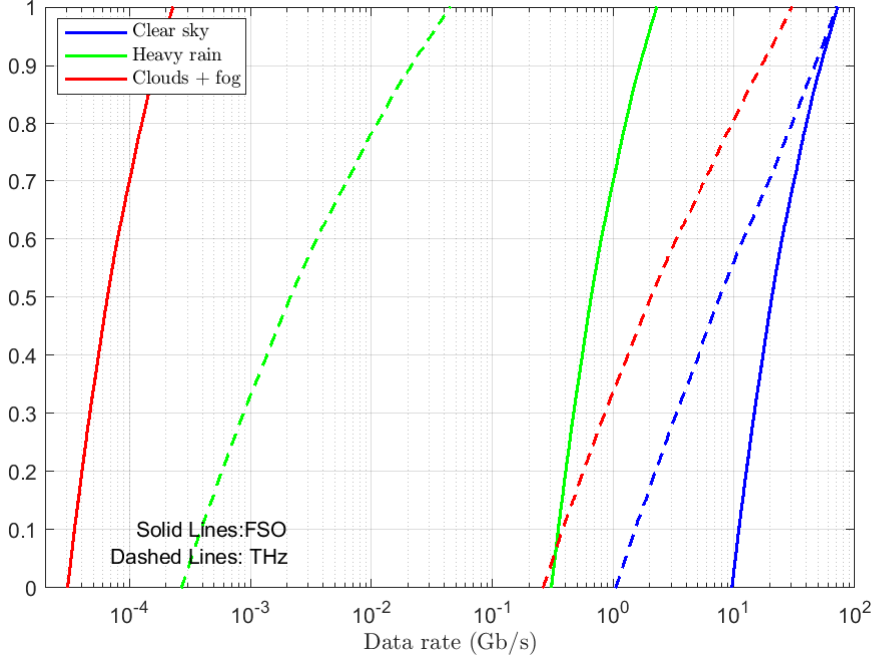}
    \caption{Cumulative Distribution Function (CDF) of the data rates achieved at the terrestrial BS and UAVs by hybrid FSO/THz fronthaul/backhaul links.}
    \label{fig:Backhaul2}
    \vspace{-12pt}
\end{figure}

The fronthaul/backhaul capabilities between the end-point HAPS and UAVs, mobile communication terminals, and ground \glspl{BS} deployed in the disaster-affected region were investigated using hybrid FSO/THz communication links in Fig. \ref{fig:Backhaul2}.  The performance comparisons are given for randomly distributed UAVs and mobile terminals in an area with a $50$ km radius under different atmospheric conditions, including clear sky, clouds, fog, and rain. 
As illustrated in Fig.~\ref{fig:Backhaul2}, the FSO link exhibits superior performance to the THz link under clear sky conditions. However, clouds and fog impose significant attenuation on the vertical FSO link, mainly due to Mie scattering. In contrast, the vertical THz link performs better in cloudy and foggy conditions. Regarding rain conditions, the THz link experiences severe degradation due to strong absorption and scattering of THz signals, which result from their resonant interaction with electromagnetic waves in this frequency range. Although the FSO link is also affected by heavy attenuation in rainy weather, it performs better than the THz link under such conditions. In summary, our simulation results indicate that a hybrid FSO/THz communication approach can provide reliable large-scale disaster area connectivity, effectively adapting to different atmospheric conditions.

\begin{figure}[]
    \centering
    \includegraphics[width=.9\linewidth]{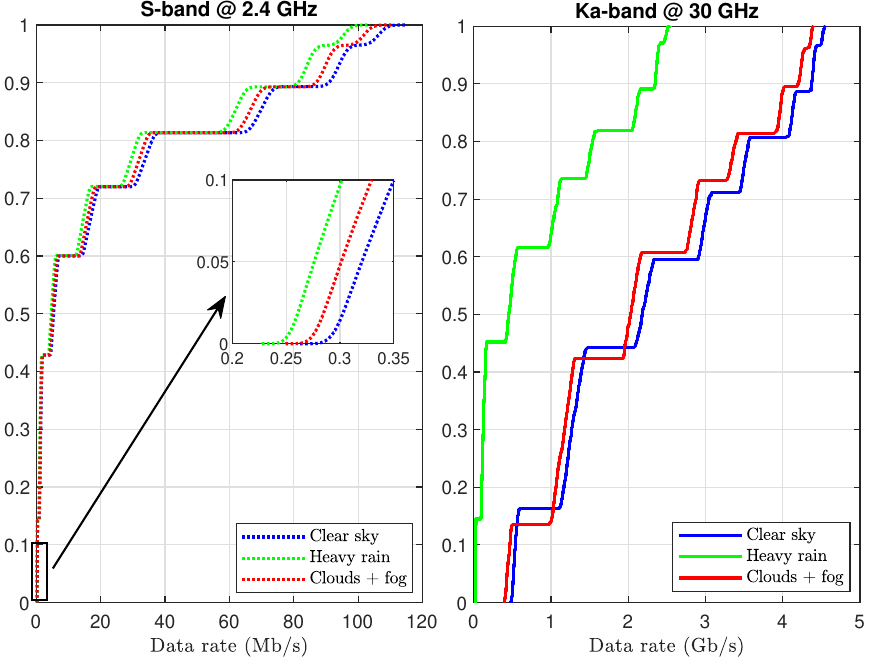}
    \caption{CDF of data rate in direct access to ground users. In the S-band, ground users are mobile handheld devices while in the Ka-band ground receivers have VSAT.}
    \label{fig:Ka_S_band_RAN}
    \vspace{-12pt}
\end{figure}

In Fig.~\ref{fig:Ka_S_band_RAN}, the \ac{RAN} connection between the end-point HAPS and the users on the ground operates in both S-band and Ka-band. S-band supports mobile handheld devices, while Ka-band requires \ac{VSAT} terminals ($60$ cm circular polarization ), which can be permanently installed or mounted on mobile platforms such as vehicles (e.g. emergency responders) or UAVs. For S-band simulations, considering an urban scenario, $1$ million users are randomly distributed within an area of $50$ km radius and served by the end-point HAPS. In the Ka-band simulations, $1000$ \ac{VSAT} devices are randomly distributed within the same area and served. All simulations incorporate HAPS parameters, air-to-ground channel characteristics, and receiver specifications as referenced in \cite{karaman2024enhancing}. The simulation results indicate that a single HAPS can replace failed ground \glspl{BS} to cover all mobile users in the area. Even in the worst case, mobile users can achieve $250$ Kbps data that enables them to receive emergency-related data. This can be broadcast information (e.g., shelter, first-aid, food locations) or precautions shared via mobile applications, as discussed in detail in the previous section. As we can see from Fig. \ref{fig:Ka_S_band_RAN}, direct access to the ground mobile users can increase up to $120$ Mbps, and this proves the proposed on-demand HAPS network can even replace commercial communication beyond public safety.  On the other hand,  \ac{VSAT} receiver antenna gain compensates for the higher pathloss in Ka-band and enhances Quality-of-Service (QoS). This allows uninterrupted high-rate communication for first responders as visioned. Moreover, the Ka-band is less affected by atmospheric conditions compared to the FSO and THz bands, whereas the S-band remains more resilient under rainy and cloudy conditions.

% Fig. \ref{fig:Backhaul} shows ...

% Fig. \ref{fig:RAN} show ...

% \hline
%         Atmospheric attenuations due to Mie scattering and Geometrical scattering & As proposed in \cite{liang2022link} & Cloud, fog, and rain attenuations & Based on \cite{alzenad2018fso} \\
%         \hline
%         Transmitter and Receiver antenna gains & As proposed in \cite{liang2022link} & - & - \\
%         \hline
%         Cloud, fog, and rain attenuations & As proposed in \cite{alzenad2018fso} & - & - \\

%\subsection{Different Scenarios}

\section{Standardization Efforts and Current Initiatives}
%\subsection{Standardization and Protocols for HAPS and NTN-Based Public Safety Communication} 

In the previous section, we demonstrated that the proposed HAPS-assisted communication network can play a crucial role in disaster response efforts. However, integrating HAPS public safety communication requires adherence to specific international standards, frameworks, and protocols. These standards define how HAPS, or other Non-Terrestrial Networks (NTN) such as satellites and terrestrial networks work together to provide resilient, scalable, and low-latency emergency communication systems. This section provides an overview of key global standards and protocols that define the role of HAPS and NTN in public safety communication.

\subsection{3GPP Standardization for NTN and HAPS Integration}

The 3rd Generation Partnership Project (3GPP) serves as the primary standardization body responsible for defining how NTN integrate with terrestrial \ac{5G} and beyond. The framework includes HAPS, Low Earth Orbit (LEO), Medium Earth Orbit (MEO), and Geostationary Orbit (GEO) satellites. A major advancement in this domain is 3GPP Release 17, which introduces NTN-specific enhancements to 5G networks. Key improvements include HAPS-to-ground and HAPS-to-satellite communication, NTN integration with 5G New Radio (5G NR), and support for Direct-to-Device (D2D) NTN communication.  Building upon this foundation, 3GPP Release 18 expands NTN capabilities by incorporating AI-driven network optimization for public safety applications, enhancing HAPS-satellite-terrestrial interoperability. The ongoing evolution of 3GPP standards will further refine the integration of HAPS into NTN architectures, ensuring seamless connectivity in emergency scenarios.

\subsection{ITU-R Standards and Regulations for HAPS-Based Communication}

The International Telecommunication Union Radiocommunication Sector (ITU-R) defines HAPS spectrum allocation, network interoperability, and disaster management in the Radio Regulations (RR). Operating at altitudes between 20-50 km, HAPS platforms provide fixed broadband access and emergency connectivity, especially in disaster-stricken and underserved regions. ITU-R F.1500 specifies technical characteristics and spectrum allocations for HAPS in the 47/48 GHz and 2 GHz bands, ensuring compatibility with terrestrial and satellite networks while minimizing interference. ITU studies estimate spectrum needs ranging from 396 MHz to 2,969 MHz for ground-to-HAPS links and 324 MHz to 1,505 MHz for HAPS-to-ground, covering broadband and disaster relief applications. At the ITU World Radiocommunication Conference 2023, HAPS was formally recognized as a High-Altitude IMT Base Station (HIBS), enabling aerial 5G broadband services. Spectrum identification in the 2 GHz and 2.6 GHz bands allows HAPS to deliver real-time emergency communication, network restoration, and mobile broadband expansion \cite{ITU_WRC23}. Continued ITU-R standardization will be key to ensuring HAPS scalability, resilience, and integration into future 5G and NTN ecosystems.  

\subsection{Efforts from Industry for HAPS-Based Communication}

In addition to government and international regulatory efforts, industry-led standardization initiatives play a critical role in defining the operational framework for HAPS-enabled communications networks. The HAPS Alliance, a global consortium that includes SoftBank, Airbus, and Aerostar, is actively working to develop technical standards for air-to-ground, air-to-satellite, and inter-HAPS communications \cite{HAPS_Regulatory_2024}. Research efforts within the alliance focus on optimizing HAPS network topology, spectrum management, and system interoperability to ensure that these platforms can be rapidly deployed for public safety and emergency response operations. One of the most significant real-world deployments of HAPS for public safety was Google's Loon project, which demonstrated the feasibility of HAPS-powered emergency communications during Hurricane Maria in Puerto Rico (2017). Following the disaster, Loon deployed its stratospheric balloons and restored mobile network connectivity to over 200,000 people in the affected region. This highlighted the role of HAPS-based emergency response and demonstrated the restoration of low-latency and high-reliability networks. Similarly, in Asia, Japanese telecommunications companies, including NTT, KDDI, SoftBank, and Rakuten Mobile, have collaborated on disaster response initiatives, conducting joint training exercises with marine vessels and shared refueling stations to ensure rapid network restoration during large-scale natural disasters \cite{NTT_Disaster_Response_2025}.

% \cite{Vittori_HAPS_Thesis_2024}.
%Furthermore, academic research and industry collaboration continue to push HAPS technology forward, focusing on persistent stratospheric platforms, AI-driven network management, and NTN-based 6G integration. Advances in solar energy, lightweight materials, and low-power communication systems will further enhance the capabilities of HAPs in disaster response, global connectivity, and environmental monitoring \cite{9782095}.

\subsection{Space Agency Initiatives for HAPS-NTN Public Safety}

Various national space agencies and research institutions are also contributing to developing HAPS-based NTN public safety networks.  In the United States, NASA is leading efforts in HAPS-NTN emergency communications in collaboration with NOAA, FEMA, and the U.S. Department of Defense. The NASA HAPS-NTN Emergency Communications Study examines the role of stratospheric communication platforms for wildfire, hurricane, and earthquake response. This research includes AI-driven predictive modeling and optimizing the placement of HAPS networks for disaster scenarios. The Strategic Tactical Radio and Tactical Overwatch (STRATO) Project by NASA explores LTE-based HAPS connectivity for emergency responders, enhancing situational awareness in remote fire-prone regions \cite{NASA_HAPS_2023}. The European Space Agency (ESA) is leading the SAT-HAPS Convergence Initiative, integrating HAPS and satellite-based emergency communication networks into European disaster management frameworks. This initiative aims to strengthen multi-layered NTN architectures, ensuring continuous public safety communication even in disaster-affected zones \cite{fehr2018high}.% In Asia, Japanese telecommunications companies, including NTT, KDDI, SoftBank, and Rakuten Mobile, have collaborated on disaster response initiatives, conducting joint training exercises with marine vessels and shared refueling stations to ensure rapid network restoration during large-scale natural disasters \cite{NTT_Disaster_Response_2025}. %Meanwhile, China is exploring Beidou-HAPS integration, integrating HAPS-assisted emergency communication networks using Beidou satellite navigation \cite{10421558}.

%Additionally, SoftBank is advancing HAPS technology to maintain communication services during disasters, ensuring connectivity even when terrestrial base stations fail due to earthquakes or tsunamis \cite{ SoftBank_HAPS_2024}.

\subsection{Future Directions in HAPS and NTN Public Safety Standards}

As the role of HAPS in public safety communications continues to grow, future standardization efforts will focus on the full integration of HAPS into next-generation NTN architectures. This includes refining the HAPS to 5G NTN protocols in 3GPP Release 18 and beyond, extending ITU-R frequency allocations to support global HAPS deployment, and developing AI-driven network optimization models to improve real-time emergency communications performance. In addition, the emergence of multi-layer HAPS satellite-terrestrial architectures will improve the resilience and efficiency of emergency communications networks. The combination of HAPS, LEO, MEO, and GEO satellite constellations will enable seamless connectivity in disaster-prone, remote, and underserved regions, ensuring that first responders and affected populations have access to reliable communication channels during emergencies.

\section{Conclusions}

Reliable communication is essential for effective disaster relief, but traditional infrastructures often fail in large-scale crises. While solutions like cloud-based systems and portable units offer some resilience, they remain limited when both radio access and backhaul networks are disrupted. This paper has explored the potential of HAPS as a transformative solution for disaster communications. 
%Unlike traditional systems, HAPS can not only provide on-demand connectivity for mobile users, but also restore backhaul connections, ensuring seamless communication between affected areas and disaster management centers. 
In contrast to conventional systems, HAPS can reestablish backhaul connections in addition to offering mobile users on-demand access, guaranteeing smooth communication between disaster management centers and impacted areas.
By acting as an aerial bridge, HAPS facilitates real-time data collection from disaster-affected regions, provides critical information to emergency responders, and improves situational awareness for better coordination of relief efforts. The simulation results show that integrating HAPS with hybrid FSO/THz communication links enables the achievement of sufficient backhaul capacity and network resilience, even under adverse atmospheric conditions. In addition, HAPS-enabled RAN connectivity in S- and Ka-band ensures robust communications for first responders and the affected population, which can be adapted to various disaster scenarios. Overall, HAPS-assisted networks offer a promising pathway toward a more resilient and adaptive emergency communication framework. Future research should prioritize optimizing deployment strategies, enhancing energy efficiency, and integrating AI for dynamic network management. By harnessing these advancements, HAPS can play a pivotal role in minimizing communication disruptions and enhancing disaster response effectiveness, ultimately saving lives and reducing the overall impact of disasters.

%This paper highlights the potential of HAPS as a rapidly deployable, resilient communication solution. Beyond providing connectivity for mobile users, HAPS restores backhaul links, bridges disaster management centers with affected areas, and facilitates real-time data exchange. 

%Simulation results confirm that integrating HAPS with hybrid FSO/THz links enhances network resilience, while S-band and Ka-band connectivity ensures reliable communication for responders and affected populations. HAPS-assisted networks offer a promising path toward more adaptive disaster communication. Future research should focus on optimizing deployment, energy efficiency, and AI-driven network management to maximize their effectiveness in emergency scenarios.

% Can use something like this to put references on a page
% by themselves when using endfloat and the captionsoff option.

%\vspace{-10pt}
\bibliographystyle{IEEEtran}
\bibliography{bibliography}

\end{document}